\documentclass[conference]{IEEEtran}
\usepackage{cite}
\usepackage{amsmath,amssymb,amsfonts}
\usepackage{algorithmic}
\usepackage{graphicx}
\usepackage{textcomp}
\usepackage{xcolor}

\usepackage[shortcuts,acronym]{glossaries}
\newacronym{API}{API}{application programming interface}
\newacronym{SOA}{SOA}{state-of-the-art}
\newacronym{ANNS}{ANNS}{approximate nearest neighbor search}
\newacronym{IaC}{IaC}{Infrastructure as Code}
\newacronym{CI}{CI}{cyberinfratructure}
\newacronym{CICD}{CI/CD}{continuous integration and continuous delivery}
\newacronym{ETS}{ETS}{energy-to-solution}
\newacronym{NDN}{NDN}{named data networking}
\newacronym{ICN}{ICN}{information-centric networking}
\newacronym{OSI}{OSI}{Open Systems Interconnection}
\newacronym{DSI}{DSI}{data-sharing interface}
\newacronym[plural=DAGs, firstplural=directed acyclic graphs]{DAG}{DAG}{directed acyclic graph}
\newacronym{HPC}{HPC}{high-performance computing}
\newacronym[plural=WANs, firstplural=wide-area networks]{WAN}{WAN}{wide-area network}
\newacronym{PS}{Pub/Sub}{publish-subscribe}
\newacronym{TCPIP}{TCP/IP}{Transmission Control Protocol/Internet Protocol}
\newacronym{DDS}{DDS}{Data Distribution Service}
\newacronym{QOS}{QoS}{quality of service}

\usepackage{graphicx}
\usepackage{subfig}
\usepackage{float}
\usepackage[export]{adjustbox}

\def\BibTeX{{\rm B\kern-.05em{\sc i\kern-.025em b}\kern-.08em
    T\kern-.1667em\lower.7ex\hbox{E}\kern-.125emX}}
\begin{document}

\title{Accelerating the Operation of Complex Workflows through Standard Data Interfaces
}

\author{\IEEEauthorblockN{1\textsuperscript{st} Taylor Paul}
\IEEEauthorblockA{\textit{Computer Science} \\
\textit{University of Maryland}\\
College Park, USA \\
thpaul@umd.edu}
\and
\IEEEauthorblockN{2\textsuperscript{st} William Regli}
\IEEEauthorblockA{\textit{Computer Science} \\
\textit{University of Maryland} \\
College Park, USA \\
regli@umd.edu} }

\maketitle

\begin{abstract}
In this position paper we argue for standardizing how we share and process data in scientific workflows at the \textit{network-level} to maximize step re-use and workflow portability across platforms and networks in pursuit of a foundational workflow stack.  
We look to evolve workflows from steps connected point-to-point in a \ac{DAG} to steps connected via shared channels in a message system implemented as a network service.
To start this evolution, we contribute: a preliminary reference model, architecture, and open tools to implement the architecture today.
Our goal stands to improve the deployment and operation of complex workflows by decoupling data sharing and data processing in workflow steps.
We seek the workflow community's input on this approach's merit, related research to explore and initial requirements from the workflows community to inform future research.
\end{abstract}


\section{Preliminary Reference Model and Architecture}

\begin{figure}[ht]
  
    \subfloat[As-is workflow directed acyclic graph representation]{\label{fig:workflows-as-is}%
        \includegraphics[width=\linewidth, valign=t]{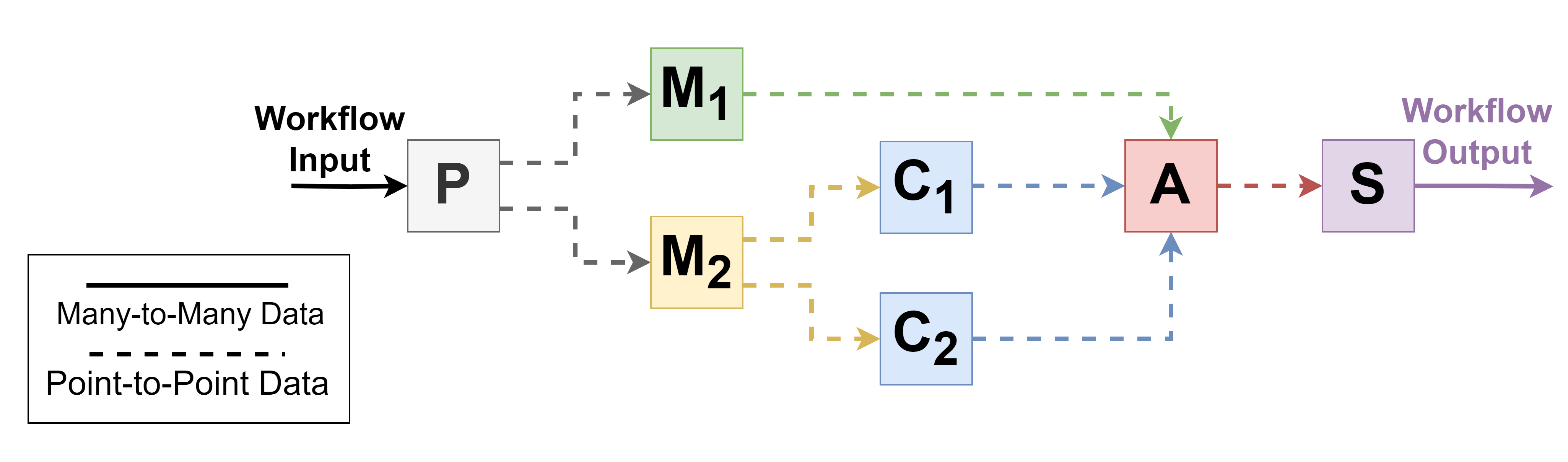}%
    }
    
    \subfloat[To-be workflow message bus representation]{\label{fig:workflows-to-be}%
        \includegraphics[width=\linewidth, valign=t]{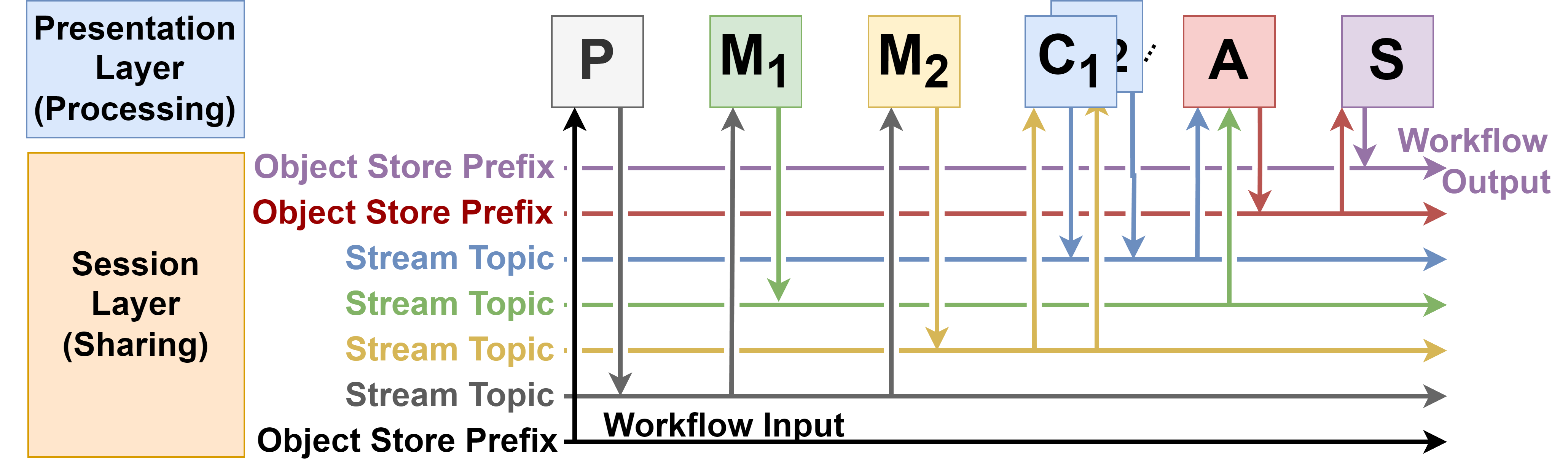}%
    }
    \caption{A depiction of the current representation of a workflow in a \ac{DAG} and the to-be representation separating processing and data sharing for independent scaling and scheduling.}
    \label{fig:workflows-as-is-to-be}
\end{figure}

Here we introduce our reference model and architecture and discuss the implications for workflows if we identify a foundational workflow stack\cite{silva2023wcs22} in each \ac{WAN} a workflow spans. 
Starting with changes at the workflow level, we propose a shift from modeling workflows strictly as \acp{DAG} (Fig. \ref{fig:workflows-as-is}) to compute components connected to a messaging system\cite{hohpe2012eip} which may span multiple \acp{WAN} (Fig. \ref{fig:workflows-to-be}). 
Conceptually, the \ac{DAG} remains a valid planning model, but now edges in the graph require a many-to-many data channel, or \ac{PS} model, instead of a point-to-point \ac{TCPIP} connection. 
This channel permits independent scaling of bottleneck steps in  a workflow (i.e. $C$ in Fig. \ref{fig:workflows-as-is}).
It also permits effective modification of workflows. If we must update step $A$ with $A_2$ in Fig. \ref{fig:workflows-to-be}, we first bring $A_2$ online, have it read $M_1$, $C_1$ and $C_2$ outputs from the shared channels and write the new output to a new object store prefix.
Once sufficient testing passes on live data, a workflow engineer can then migrate downstream steps to $A_2$'s object store prefix.

\begin{figure}[htp]
    \centering
    \subfloat[Model]{\label{fig:ref-model}%
        \includegraphics[width=0.33\linewidth, valign=b]{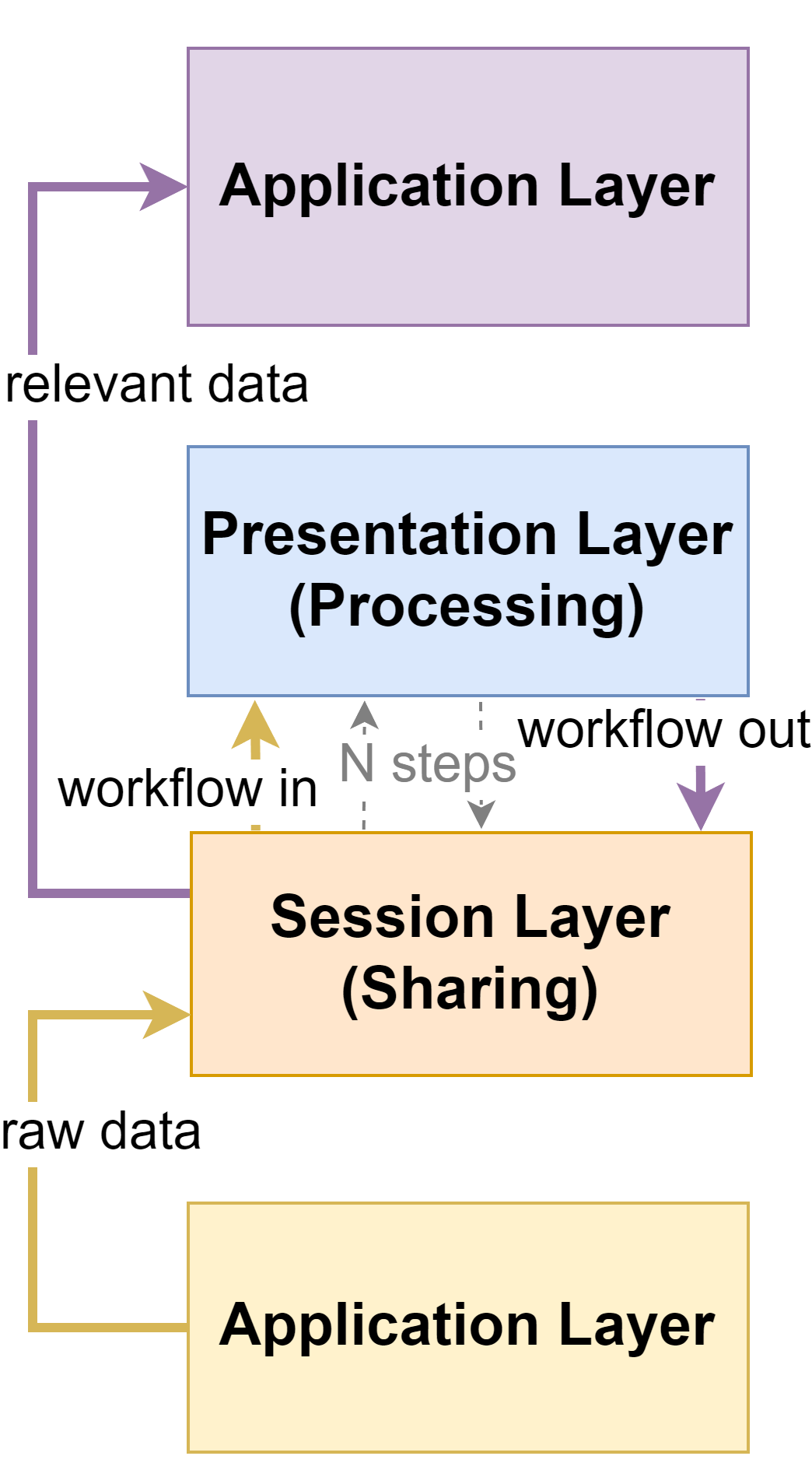}%
        \vphantom{\includegraphics[width=0.01\linewidth, valign=b]{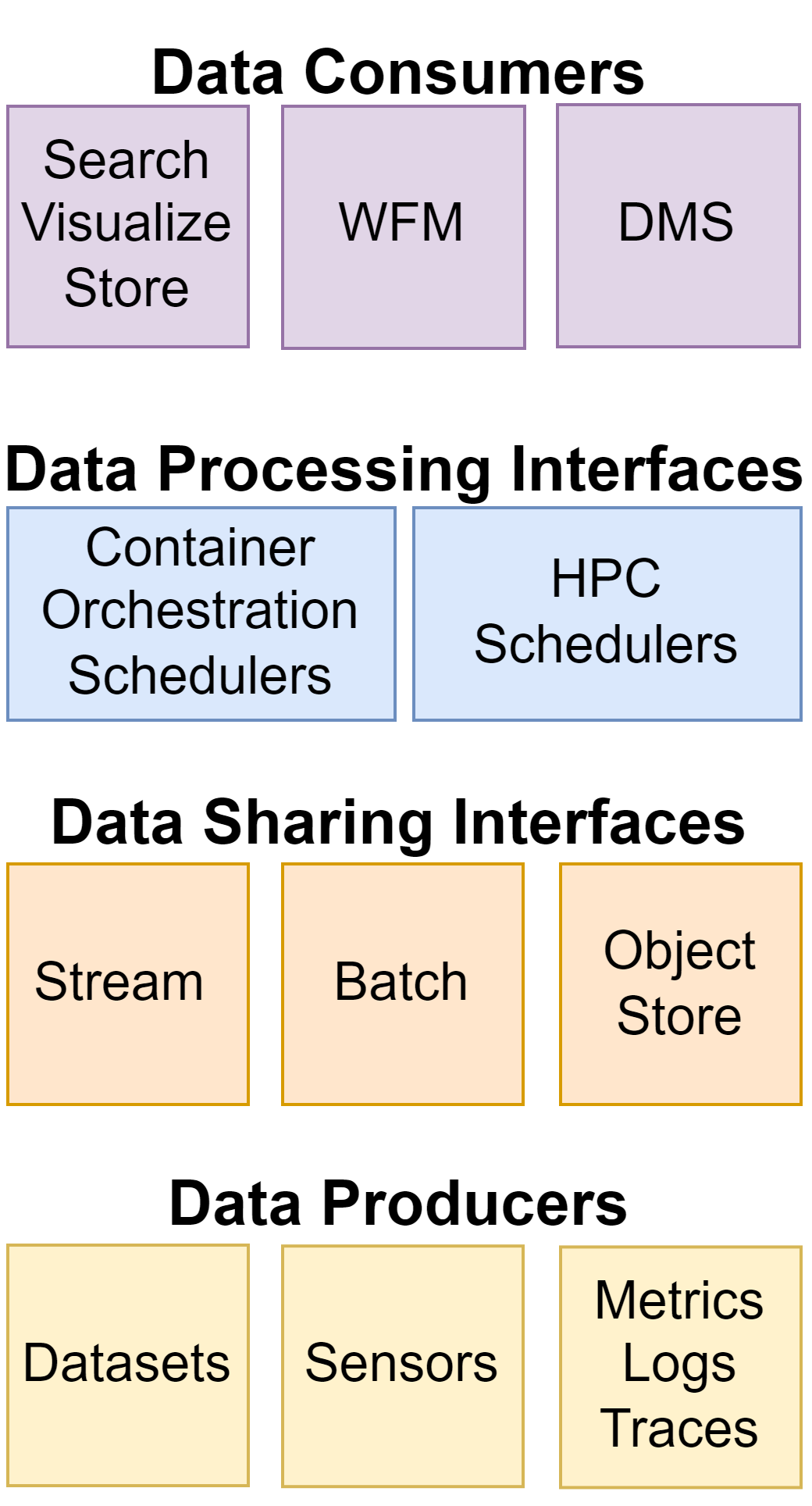}}%
    }
    \subfloat[Architecture]{\label{fig:ref-arch}%
        \includegraphics[width=0.33\linewidth, valign=b]{workflows-ref-architecture}%
    }
    \subfloat[Implementation]{\label{fig:ref-implementation}%
        \includegraphics[width=0.33\linewidth, valign=b]{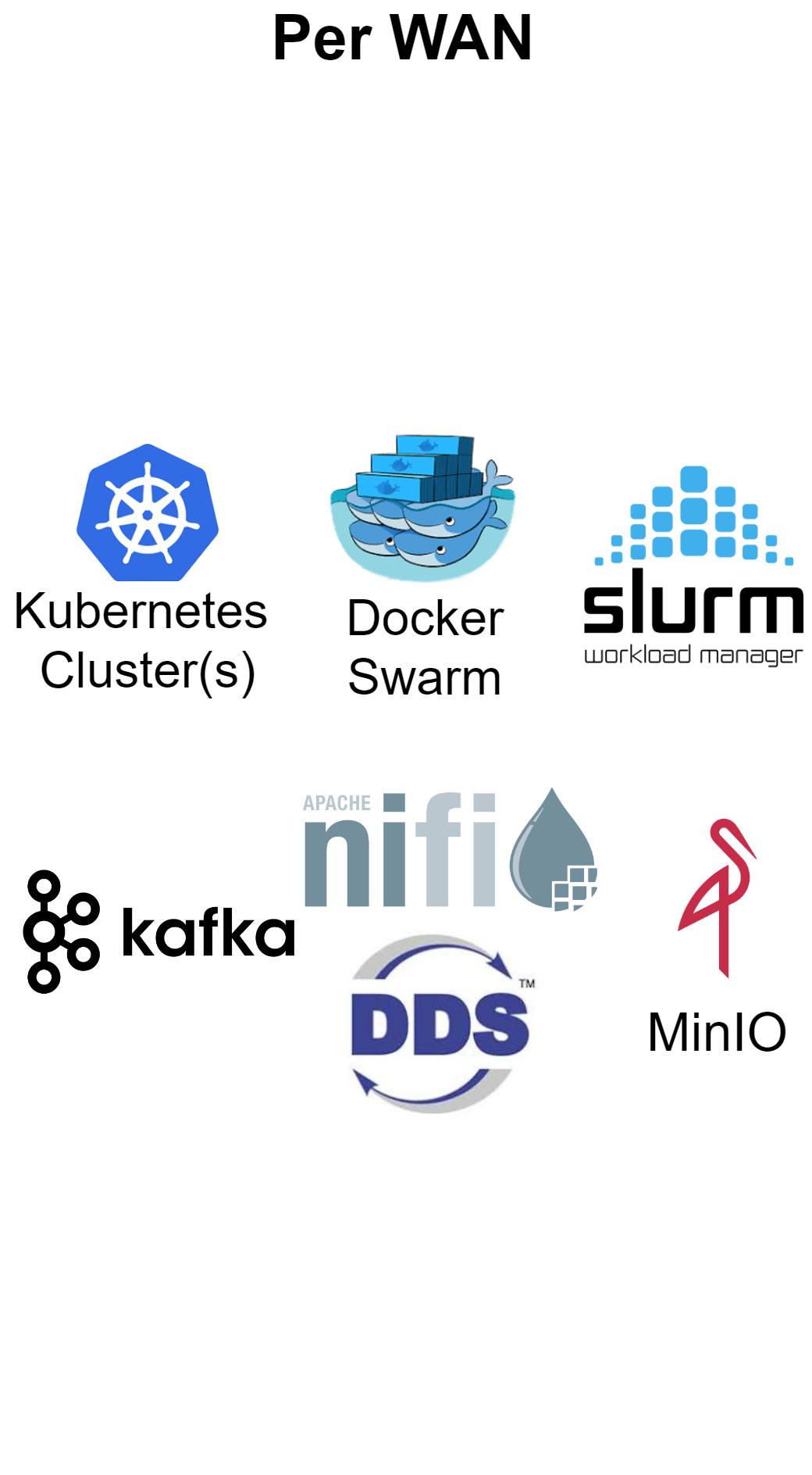}%
    }
    \caption{Proposed reference model, architecture and potential means to implement data-sharing and processing interfaces.}
    \label{fig:ref-all}
\end{figure}

We now discuss network-level changes to achieve the aforementioned workflow flexibility. 
While most large organizations endorse the \ac{OSI} model for network interoperability, in reality, the internet leverages the four-layer \ac{TCPIP} model\cite{russell2013internet}. 
Our reference model (Fig. \ref{fig:ref-model}) suggests that complex workflows mandate open protocols and hardware implementations of the two \ac{OSI} layers stuffed into applications in the \ac{TCPIP} stack: the \textit{Session} and \textit{Presentation} Layers. 
We agree with the position of others\cite{dds,hohpe2012eip} that applications, or steps in a workflow, exchanging data should avoid direct sharing in favor of middleware to achieve portability, interoperability and other desirable qualities. Our model reflects this as data producers (applications) send data to the Session Layer as workflow input, then workflow steps pull inputs and push outputs to this layer until a data consumer (application) pulls the workflow output from the Session Layer for ingest (see Fig. \ref{fig:ref-model}, \ref{fig:ref-arch} and \ref{fig:workflows-to-be}). Below the Session Layer, \ac{TCPIP} Transport, Internet and Link Layer protocols suffice.

Available open-\textit{decentralized} (peer-to-peer) middleware solutions\cite{dds} warrant experimentation to provide the Session Layer service, as do various open-\textit{centralized} solutions (e.g. Kafka, NiFi, MinIO, etc.). 
We see use cases for both, just as ad-hoc (decentralized) networks and router-based (centralized) networks exist to support network-layer connectivity. 
Ideally, one extends existing protocols\cite{dds,ddsirtps} in the future to dynamically shift from peer-to-peer to a centralized service based on data \ac{QOS} requirements, live network conditions and availability of the centralized service.

As depicted in Fig. \ref{fig:ref-implementation}, we suggest data-intensive workflows require Session-Layer devices in each network, a centralized data-distribution approach, to ease resources required from workflow input applications and individual workflow steps for data sharing downstream. We call these devices \textit{data-sharing interfaces} and they are akin to routers for the Internet Layer. They decouple workflow steps from each other; only requiring a data producer to use resources to share data with the interface, not all downstream steps. 
We can also implement access controls for all data and enforce other data standards at these interfaces.
We posit a mature architecture mandates categories of data-sharing interfaces with different performance characteristics, or \ac{QOS}, in terms of max message size, latency and throughput. 
For example, exchanges of small messages requiring sub-second delay should employ a \textit{stream interface} implemented in a stand-alone software system (e.g. Kafka, others). Data exchanges of large, infrequently-updated files certainly do not require the same \ac{QOS} nor should we force them into a data stream. Therefore, a \textit{object store interface} should support reading and writing of such files among workflow steps. 
We see both the stream and object store interfaces implementing a \ac{PS} model. 
For the \textit{batch interface}, we envision a tool like Apache NiFi pulling data from steps running in legacy systems via a protocol that application accepts (e.g. SFTP) and pushing to a stream or object store to serve downstream steps. 
Also, we see a batch interface windowing streams of small events into batches written to object storage, perhaps to efficiently feed a machine-learning algorithm's inference or training steps. 
Initially, we expect these interfaces implemented in \acp{WAN} supporting complex workflows in an \`a la carte manner; only those required. 

Leveraging Session-Layer interfaces, in the Presentation Layer we free workflow steps from \ac{WAN}-specific-data dependencies and through containerization from external software dependencies.
The original \ac{OSI} paper \cite{zimmermann1980osi} states: 
\begin{quote}
The  purpose  of  the  Presentation  Layer is to provide the set of services which  may be selected  by  the Application Layer to enable it  to  interpret  the meaning  of the data exchanged.
\end{quote}
We argue that most workflow steps fit into this definition. 
Machine-learning algorithms too as they process data to enrich (i.e. classification, score, embeddings) or summarize for applications to display.
This view frees most workflow steps from specific servers enabling movement across networks.

Now we envision an architecture with \textit{data-processing interfaces} that permit scheduling and running of containerized workflow steps, or \ac{HPC} compute steps, sharing data through our data-sharing interfaces. 
Since we standardize data sharing across \acp{WAN} for inputs and outputs, we more easily move, update, inject or delete \textit{decoupled} steps in distributed workflows with minimal changes required across the entire workflow.

\section{Conclusion}

We presented a network-level approach to identifying the foundational workflow stack with recommendations for standard data processing and sharing services via consistent interfaces across \acp{WAN}. This approach differs from industry and academic focus on improving data operations within proprietary and/or open platforms at the Application Layer; a result of the \ac{TCPIP} model's dominance. If these interfaces interest the workflow community, future work exists: 1) empirical testing of data-sharing interfaces in workflows, 2) supporting efficient data replication, as required, between interfaces in separate \acp{WAN} with metrics collection, 3) profiling steps for realistic compute requests, 4) designing a scheduler to distribute workflow steps\cite{Lehmann2023sched, skluzacek2024towards} and data at the optimal interfaces across  \acp{WAN}. 
We argue that implementing these solutions at the right layers in the network will accelerate the deployment and operation of future workflows.

\bibliographystyle{IEEEtran.bst}
\bibliography{IEEEabrv, benchmark_readings}
\end{document}